\documentclass[11pt]{article}

\usepackage[preprint]{acl}

\usepackage{times}
\usepackage{latexsym}

\usepackage[T1]{fontenc}
\usepackage[utf8]{inputenc}

\usepackage{microtype}

\usepackage{inconsolata}

\usepackage{enumitem}
\usepackage{latexsym}
\usepackage{fontawesome}
\usepackage{pifont}
\usepackage{listings}[language=Python]
\usepackage{graphicx}
\usepackage{booktabs}
\usepackage{mdframed}
\usepackage{lipsum}
\usepackage{placeins}
\usepackage{multirow}
\usepackage{amsmath}
\usepackage{hyperref}
\usepackage{cancel}
\usepackage{array}
\usepackage[subtle]{savetrees}
\usepackage{xspace}
\usepackage{xcolor}
\usepackage{algorithm}
\usepackage[noend]{algpseudocode}
\usepackage{amssymb}

\definecolor{mygray}{gray}{0.97}
\colorlet{shadecolor}{mygray}

\newmdenv[%
  backgroundcolor=mygray, % Set the background color
  linewidth=0pt
]{newshaded}

\definecolor{codegreen}{rgb}{0,0.6,0}
\definecolor{codegray}{rgb}{0.5,0.5,0.5}
\definecolor{codepurple}{rgb}{0.58,0,0.82}
\definecolor{backcolour}{rgb}{0.95,0.95,0.92}

\definecolor{ForestGreen}{HTML}{029147}

\lstdefinestyle{mystyle}{
    language=Python,
    backgroundcolor=\color{backcolour},   
    commentstyle=\color{codegreen},
    keywordstyle=\color{blue},
    numberstyle=\tiny\color{codegray},
    stringstyle=\color{codepurple},
    basicstyle=\ttfamily\footnotesize,
    breakatwhitespace=false,         
    breaklines=true,                 
    captionpos=b,                    
    keepspaces=true,                 
    numbers=left,                    
    numbersep=5pt,                  
    showspaces=false,                
    showstringspaces=false,
    showtabs=false,                  
    tabsize=2,
    columns=fullflexible,
}

\DeclareMathOperator*{\argmax}{arg\,max}

\title{McMining: Automated Discovery of Misconceptions in Student Code}

\author{\textbf{Erfan Al-Hossami} \and \textbf{Razvan Bunescu} \\
       University of North Carolina at Charlotte\\
        Charlotte, NC, USA\\
        \texttt{\{ealhossa, rbunescu\}@charlotte.edu}}

\begin{document}
\maketitle
\begin{abstract}
When learning to code, students often develop misconceptions about various programming language concepts. These can not only lead to bugs or inefficient code, but also slow down the learning of related concepts.
% As such, it is important that misconceptions are identified and fixed as early as possible. 
In this paper, we introduce McMining, the task of mining programming misconceptions from samples of code from a student. To enable the training and evaluation of McMining systems, we develop an extensible benchmark dataset of misconceptions together with a large set of code samples where these misconceptions are manifested. We then introduce two LLM-based McMiner approaches and through extensive evaluations show that models from the Gemini, Claude, and GPT families are effective at discovering misconceptions in student code.\\
{\small
\faicon{github} \hspace{0.1em} \href{https://github.com/taisazero/mcminer}{https://github.com/taisazero/mcminer}}
\end{abstract}

\section{Introduction and Motivation}

A misconception is a belief in a false statement, in short a false belief, such as believing that the Earth is flat or that real numbers are countable. In the CS domain, programming misconceptions are beliefs about programming concepts that are not warranted by the programming language definition. We include here beliefs about the syntax and semantics of programming language constructs and builtin functions included with the language. For example, a common misconception in Python is that the {\tt range(n)} function produces integers starting at 1 and ending at $n$, which may potentially be the cause for the bug shown in Figure~\ref{fig:pc-example}.

\begin{figure}[t]
    \centering
\begin{newshaded}
\begin{minipage}{.5\columnwidth}
\small
\noindent \ding{227} {\bf Problem description}: \\

\vspace{-0.75em}
Write the {\tt factorial(n)} function that computes the factorial {\tt n!} defined as:
\begin{verbatim}
   0! = 1
   n! = n x (n - 1)!
\end{verbatim}
\noindent If the input {\tt n} is negative, the function should return {\tt 0}.

\end{minipage}\hfill
\begin{minipage}{.4\columnwidth}
\small
\vspace{1em}
\noindent \ding{227} {\bf Student code}
\begin{lstlisting}[basicstyle=\small]
1. def factorial(n):
2.   if n < 0:
3.     return 0
4.  fact = 1
5.  for i in range(n):
6.    fact = fact * i 
7.  return fact
\end{lstlisting}
\end{minipage}
\vspace{.75em}
\begin{small}

\noindent \ding{227} {\bf Potential misconception}:

\vspace{0.25em}
\noindent {\tt range(n)} produces values from 1 to {\tt n} inclusive.
\end{small}

\end{newshaded}
    \caption{Problem-code pair that exhibits a {\it potential} programming misconception about the {\tt range} function.}
    \label{fig:pc-example}
\end{figure}

Programming misconceptions are often the cause of bugs, which is obviously detrimental. Furthermore, in an educational context, student learning is impeded significantly when misconceptions slow down their ability to solve problems correctly, e.g. writing code that passes test cases. Misconceptions, knowledge gaps, and other types of difficulties~\cite{qian_students_2017} encountered by students make it difficult to learn one concept correctly, which then makes it harder to acquire other closely linked concepts, propagating and compounding throughout a course in a snowballing manner ~\cite{robins_learning_2010}. To maintain positive learning momentum, it is therefore essential that misconceptions are identified and fixed as early as possible. Typically, identifying a misconception is done by the students themselves, alone or ideally under Socratic guidance through conversations with an instructor or teaching assistant~\cite{erfan_2024_sigcse}. However, this process can be cognitively demanding, and, due to insufficient TA resources~\cite{yadav2016expanding}, not fast enough to keep up with the volume of concepts introduced in a course.

Recognizing the importance of identifying misconceptions early, in this paper we introduce {\sc McMining}, the novel task of mining programming misconceptions from code samples produced by a student over time. Correspondingly, we describe the development of a benchmark dataset of programming misconceptions together with a large set of problem descriptions and corresponding code samples that exhibit these misconceptions (Section~\ref{sec:definition}). We then introduce two versions of an LLM-based tool for mining misconceptions, {\sc McMiner-S} that identifies potential misconceptions in one code sample at a time, and {\sc McMiner-M} which attempts to identify misconceptions as patterns exhibited by multiple code samples from a student (Section~\ref{sec:mcminer}). We present positive results of the two version of the {\sc McMiner} tool on the benchmark dataset, using different LLMs (Section~\ref{sec:experiments}). The paper ends with conclusions and thoughts on future work (Section~\ref{sec:conclusion}).

\section{Related Work}
\label{sec:related}

Traditional automated systems~\cite{johnson1984proust,sirkia2012exploring,sidelib_2023} rely on hand-crafted rules targeting a limited set of specific misconceptions. Similarly, modern LLM-based approaches such as~\cite{lee_improving_2024} focus on classification of logical errors into a predefined set of categories. Overall, targeting only predefined or common misconceptions is a significant limitation, considering that educator beliefs about common errors can diverge significantly from actual student patterns~\cite{brown2014investigating}. In parallel, other approaches aim to identify code segments that contain logical errors, such as \cite{mens2021good, hoq_automated_2025}, {\it inter alia}. While knowing the location of logical errors is important, current approaches do not articulate what the actual misconception is about, which can be argued to be equally important from an educational standpoint. In this context, the {\sc McMining} task is novel in that it requires not only the identification of potentially unknown misconceptions from consistent samples of student code, but also articulating their description in terms of a false belief about a programming concept.

% Evaluating misconception detection systems remains challenging due to limited benchmarks. The Blackbox dataset~\cite{altadmri201537} provides 37 million compilation events from over 250,000 students, offering unprecedented scale for error pattern analysis. However, compilation errors capture only surface symptoms. Many misconceptions manifest as syntactically valid but logically incorrect code. Systematic assessment efforts include concept inventories~\cite{goldman2010setting} adapted from physics education and Chiodini et al.'s~\cite{python_misc_bank_2021} cross-language misconception catalog, though the latter lacks code samples for training.

\section{Task Definition and Benchmark Dataset}
\label{sec:definition}

Given a set of problems and the corresponding code samples produced by the student, misconception mining is the task of identifying any potential programming misconception that is exhibited in their code. More formally, the input to the misconception mining task consists of a set $PC$ of problem-code pairs $(p, c)$:
\begin{equation}
    \small
    PC = \{(p_1, c_1), (p_2, c_2), \ldots, (p_N, c_N)\} = \left\{ \left(p_n, c_n\right) \right\}_1^N
\end{equation}
where each pair $(p_n, c_n)$ contains a problem description $p_n$ and the corresponding coding solution $c_n$ provided by the student. Figure~\ref{fig:pc-example} shows an example $(p, c)$ pair, where the code illustrates a common misconception about the range function. When presented with this and other $(p,c)$ pairs that exhibit the same misconception, the model is expected to generate a natural language description of the misconception.
% such as the student believes that {\it "range(n) starts at 1 and ends at n"}.
Note that all programming misconceptions identified by the model are considered to be {\it potential} misconceptions. 
%While a misconception mined by the model is expected to reflect a pattern in the code samples provided by the student, the identified pattern does not always imply that the student truly holds that misconception. 
Looking at the example in Figure~\ref{fig:pc-example}, it is possible that the student has the correct knowledge about the {\tt range(n)} function and that the bug in their code is caused by mistakenly using {\tt i} instead of {\tt (i + 1)} on line 6. The more samples of buggy code that utilize the {\tt range} function are in the input set $PC$, the more likely it is that the student who wrote that code holds this misconception. 
% Determining whether the student truly holds a potential misconception identified by the model can be done by engaging the student in a conversation  that probes their knowledge of the programming construct associated with the misconception. We leave the implementation of such a probing step for future work.
%It is important to note that, while the student may have misconceptions about other relevant domains, such as mathematical terms used in the problem description that are misunderstood, the misconception mining task in this paper is focused solely on identifying programming misconceptions. Henceforth, for simplicity, we will use the term misconception to refer to a programming misconception. Furthermore, 

While misconceptions are known to often cause bugs, they do not necessarily do so. For example, a student may incorrectly believe that all local variables need to have a one letter name. Correspondingly, we classify misconceptions into two disjoint categories, {\it harmful} vs. {\it benign}, depending on whether they cause or not a bug in the code.
% A misconceptions is benign if it does not affect the correctness of the samples provided to the model. Conversely, a misconceptions is deemed harmful if it causes bugs in the code samples provided to the model.

% Also, local vs. global misconceptions. Talk about potential as well.
% An output misconception is considered correct if it is exhibited by at least one of the code samples provided as input. 

%\section{Benchmark Dataset}
%\label{sec:benchmark}

To enable the evaluation of misconception mining tools, we developed a benchmark dataset $\mathcal{D}$ of $K$ student submission sets, where each student submission set $PC^{(k)}$ contains problem-code pairs created such that a subset of the pairs exhibit a certain programming misconception $m^{(k)}$. More formally, the dataset contains the following components:
\begin{equation}
    \small
    \mathcal{D} = \left\{ \left( PC^{(k)}, \;\; mc^{(k)} \right) \right\}_1^K ;\quad
    PC^{(k)} = \left\{ \left(p^{(k)}_n, \;\; c^{(k)}_n\right) \right\}_1^{N_k}
\end{equation}
Each example in $\mathcal{D}$ consists of two parts: the set of $(p, c)$ samples $PC^{(k)}$ as the observed {\it input}, and the misconception $mc^{(k)}$ as the target {\it label}. 
%While, for generality, we could have theoretically set $mc^{(k)}$ to be a (possibly empty) set of misconceptions whose cardinality depends on the student $k$, in this paper we assign exactly one misconception per student. In the rest of this section, we describe how we created each component of the dataset, namely the misconceptions $mc^{(k)}$, the problems $p^{(k)}_j$, and the code samples $c^{(k)}_j$ that exhibit misconception $mc^{(k)}$.

To develop $\mathcal{D}$, we first created a set $\mathcal{MC} = \left\{ mc^{(m)}\right\}_1^M$ of $M = 67$ misconceptions, of which 64 are common, documented misconceptions, and 3 are artificial misconceptions that are included to enable evaluation on novel misconceptions. 
We then created a set $\mathcal{PS} = \left\{(p_n, S_n)\right\}_1^N$ of $N = 501$ problem-solution pairs $(p, S)$, where for each problem $p$, the set $S$ contains one or more {\it correct} coding solutions for that problem.

\begin{algorithm}[t]
\small
\caption{Benchmark Dataset Generation}
\label{alg:benchmark-generation}
\begin{algorithmic}[1]
\State Initialize dataset $\mathcal{D} \gets \emptyset$
\For{each student $k \in \{1, 2, \ldots, K\}$}
    \State Select misconception $mc^{(k)} \in \mathcal{MC}$
    \State Initialize problem-code set $PC^{(k)} \gets \emptyset$
    \State Sample from $\mathcal{PS}$ a set of $N_k$ problem-solution pairs
    \Statex \hspace{1.5em} $PS^{(k)} = \left\{ \left(p^{(k)}_n, S^{(k)}_n\right)\right\}_1^{N_K}$
    \For{each $\left(p^{(k)}_n, S^{(k)}_n\right) \in PS^{(k)}$}
        \State Sample a correct solution $s^{(k)}_n \in S^{(k)}_n$
        \State {\sc McInject} $mc^{(k)}$ in $s^{(k)}_n$ to obtain code $c_n^{(k)}$
        \State Add problem-code pair $\left(p^{(k)}_n, c_n^{(k)}\right)$ to $PC^{(k)}$
    \EndFor
    \State Add $\left(PC^{(k)}, mc^{(k)}\right)$ to dataset $\mathcal{D}$
\EndFor
\State \Return dataset $\mathcal{D}$ of $K$ misconception-labeled code sets
\end{algorithmic}
\end{algorithm}

Given $\mathcal{MC}$ and $\mathcal{PS}$, the benchmark dataset $\mathcal{D}$ is created as shown in Algorithm~\ref{alg:benchmark-generation}. An essential component of this procedure is shown at step 8, where a misconception is injected into the correct coding solution of a problem in order to create code that exhibits that misconception. In Section~\ref{sec:mcinject} below we describe the {\sc McInject} tool that we developed for this purpose.

\begin{table}[t]
\small
\centering
\begin{tabular}{lr}
\toprule
Misconceptions & 67 \\
Problem-Solution pairs used & 25 \\
%(from 501 available) & \\
\midrule
{\bf Code samples} & \textbf{1,675} \\
% {\bf Code samples} & \textbf{1,561} \\
\quad Exhibiting misconceptions & 1,063 \\
\quad Showing no misconception & 612 \\
%\quad Filtered out by LLM-as-Judge & 114 \\
\midrule
{\bf Bags of code samples} & {\bf 339}\\
\quad Bags with misconceptions & 279 \\
\quad Bags with correct code only & 60 \\
\midrule
Code samples per bag & 4--8 \\
\bottomrule
\end{tabular}
\caption{Benchmark dataset statistics.}
\label{tab:benchmark-stats}
\end{table}

Table~\ref{tab:benchmark-stats} summarizes the key statistics of our benchmark dataset. We organized these samples into 339 code sets, or bags of code samples, each containing between $N_k = 4$ to $N_k = 8$ code samples. 
% This bag structure allows us to evaluate both single-instance mining ({\sc McMiner-S}) and multi-instance mining ({\sc McMiner-M}). 
Importantly, 60 bags contain only correct code samples without any misconceptions, serving as negative examples to test the system's ability to correctly identify when no misconceptions are present.

\subsection{The McInject Tool}
\label{sec:mcinject}

The misconception injection tool {\sc McInject} takes as input the description of a computational problem $p$, a correct solution code $s$, and the description of a programming misconception $mc$. As output, it produces a code $c$ as a version of $s$ that is modified to exhibit the misconception $mc$. The code $c$ should look as if written by a student holding misconception $mc$ who tries to solve problem $p$. The tool is implemented using Claude Sonnet-4.5 with extended thinking, with a structured prompt and in-context learning examples. When a misconception cannot be meaningfully applied to a given solution code, {\sc McInject} is instructed to indicate so, rather than force inappropriate modifications. An LLM-as-judge evaluation shows that 90.3\% of the code samples successfully exhibit their target misconceptions. Note that this is a conservative estimate of {\sc McInject}'s performance, since the LLM-as-judge sometimes discards code samples that exhibit benign misconceptions, where the code is both correct and natural. Further manual evaluation of 88 code samples shows a 96.6\% agreement between LLM judgment and human assessment, confirming the reliability of the LLM-as-judge. Appendix~\ref{sec:appendix-mcinject} provides further detailed descriptions of the tool development, including the prompt, LLM hyper-parameters, and evaluation setting.

\section{The McMiner Tools}
\label{sec:mcminer}

As defined in Section~\ref{sec:definition}, in the misconception mining task the input consists of a bag of one or more problem-code pairs from a student, and the output is a potential misconceptions exhibited in the code samples. To solve this task, we developed two variants of an LLM-based mining tool: a {\it single instance} variant and a {\it multiple instance} variant.

The single instance variant {\sc McMiner-S} proceeds in two stages, as shown in Algorithm~\ref{alg:mcminer-s}. First, an LLM is instructed to identify potential misconceptions in each problem-code pair (steps 2 to 4), using the prompt shown in Appendix~\ref{sec:appendix-mcminer}. Then, out of all the identified misconceptions, the one found in the most problem-code pairs is returned (steps 5 to 10).

\begin{algorithm}[!t]
\small
\caption{{\sc McMiner-S}}
\label{alg:mcminer-s}
\textbf{Input}: Bag of pairs $PC = \{(p_1, c_1), (p_2, c_2), \ldots, (p_N, c_N)\}$
\textbf{Output}: Potential misconception $\hat{mc}$ for the bag
\begin{algorithmic}[1]
\State Initialize set of misconceptions $M \gets \emptyset$
\For{each problem-code pair $(p_j, c_j) \in PC$}
    \State Identify potential misconception $mc_j$ for $(p_j, c_j)$
    \State If found, add $mc_j$ to $M$
\EndFor
\For{each misconception $mc \in M$}
    \State Let $count(mc) = \left|\{(p_j, c_j) \in PC | mc_j = mc\}\right|$
\EndFor
\If{all counts are 0} \Comment{\it no misconception found}
    \State \Return $\epsilon$ 
\Else \Comment{\it some misconception found}
    \State \Return $\hat{mc} = \displaystyle\argmax_{mc \in M} \;\; count(mc)$
\EndIf
\end{algorithmic}
\end{algorithm}

%The multiple-instance variant, described in Section~\ref{sec:multiple}, an LLM is given as input directly the entire set of problem-code pairs and asked to identify misconceptions by looking for patterns that appear in as many code sample as possible. 
% Note that the misconception mining problem, as defined in Section~\ref{sec:definition}, is reminiscent of the Multiple Instance Learning (MIL) setting~\cite{dietterich_solving_1997}, where the machine learning model is expected to produce a label for a bag of instances. Similar to the MIL setting, we consider that a collection (bag) of program-code pairs exhibits a potential misconception $mc$ if and only if at least one program-code pair (instance) exhibits the misconception $mc$.

In the multiple instance variant  {\sc McMiner-M}, an LLM is given the entire bag of program-code pairs and is instructed to identify the misconception that is shared by the largest number of code samples in the bag, using the prompt shown in Appendix~\ref{sec:appendix-mcminer}. 
% If no misconception is found, the LLM is instructed to output the empty misconception $\epsilon$.
Compared to {\sc McMiner-S}, {\sc McMiner-M} has the advantage of looking at multiple code samples that may exhibit the same potential misconception pattern, which in theory should enable more precision in identifying misconceptions. 
% As described in Section~\ref{sec:definition}, using solely the example shown in Figure~\ref{fig:pc-example}, it is not entirely clear that the student has a misconception about the {\tt range(n)} function. However, seeing the same potential misconception in multiple, different code samples, increases the likelihood that it is indeed a misconception.

\begin{table*}[t]
\small
\centering
\begin{tabular}{|l|cccc|c|}
\cmidrule{2-6}
\multicolumn{1}{c}{} & \multicolumn{4}{|c|}{\sc McMiner-M} & \multicolumn{1}{c|}{\sc McMiner-S} \\
\midrule
\textbf{Language Model} & \textbf{Precision} & \textbf{Recall} & \textbf{F1-Score} & \textbf{Accuracy} & \textbf{Accuracy} \\
\midrule
OpenAI o3-mini (low-effort) & 85.6\% & 67.5\% & 75.5\% & 75.5\% & 76.9\% \\
OpenAI o3-mini (medium-effort) & 83.3\% & 70.8\% & 76.5\% & 76.9\% & \textbf{78.5\%} \\
\midrule
Anthropic Claude Sonnet-4.5 & 79.1\% & 77.7\% & 78.4\% & 78.7\% & 66.4\%\\
Anthropic Claude Sonnet-4.5 + Reasoning & 83.8\% & 77.2\% & \textbf{80.3\%} & \textbf{82.0\%} & 68.7\%\\
\midrule
Gemini 2.5-flash & 79.7\% & 56.2\% & 65.9\% & 61.1\% & 74.0\%\\
Gemini 2.5-flash + Reasoning & 77.0\% & 76.7\% & 76.8\% & 76.9\% & 69.4\%\\
\bottomrule
\end{tabular}
\caption{{\sc McMiner} results for multiple-instance mining (left) and single-instance mining (right).}
\label{tab:mcminer-m-results}
\end{table*}

\section{Experimental Evaluations}
\label{sec:experiments}

We used the benchmark dataset to evaluate the effectiveness of the two McMiner tools when instantiated using three LLMs: OpenAI's o3-mini models, Anthropic's Claude-Sonnet-4.5 models, and Gemini 2.5-Flash models. The experimental setup for each model is detailed in Appendix~\ref{sec:appendix-experimental}.

% Our evaluation addresses three key research questions: (1) How accurately can LLMs identify programming misconceptions from student code samples? (2) Does analyzing multiple code samples simultaneously (McMiner-M) improve misconception detection compared to single-sample mining (McMiner-S)? (3) How does the addition of reasoning capabilities affect misconception mining performance across different model families?

Table~\ref{tab:mcminer-m-results} present the performance of different LLMs in the multiple and single instance mining settings. For computing accuracy, precision, recall, F1-score, we define: {\it true positives} are cases where the Ground Truth (GT) misconception matches the prediction, or the prediction is a validated novel misconception; {\it true negatives} are cases where both GT and prediction contain no misconception; {\it false positives} are cases where the prediction does not match GT and fails validation; and {\it false negatives} are cases where GT contains a misconception but the model predicted none. Note that when a validated novel misconception is found but GT contains a different misconception, this counts as both a true positive (for the valid discovery) and as a false negative (for missing GT).

Multi-instance mining ({\sc McMiner-M}) substantially outperforms single-instance mining ({\sc McMiner-S}), with the best model (Claude Sonnet-4.5 + Reasoning) achieving 82.0\% accuracy for {\sc McMiner-M} compared to 78.5\% for the best {\sc McMiner-S} model (o3-mini medium-effort). This demonstrates the value of analyzing multiple code samples simultaneously to identify consistent patterns with higher confidence. Enabling reasoning capabilities consistently improves {\sc McMiner-M} performance across model families: Claude (78.7\% to 82.0\%), Gemini (61.1\% to 76.9\%), and o3-mini (75.5\% to 76.9\% with medium effort).  

The best {\sc McMiner-M} model achieves 93.3\% accuracy on bags with misconceptions but only 59.3\% on correct-only bags, indicating higher false positive rates on correct code.

{\sc McMiner} identifies novel misconceptions effectively: Claude Sonnet-4.5 with reasoning discovered 41 novel true positives (12.1\% of bags), including: ``The student believes that the division operator \texttt{/} and integer division operator \texttt{//} are interchangeable or that \texttt{/} will automatically return an integer when the result is a whole number.''

Misconception difficulty varies substantially: syntax errors (e.g., \texttt{=} vs. \texttt{==}) achieve $>$95\% accuracy, while subtle benign misconceptions (e.g., unnecessary parentheses) and subtle out-of-distribution cases (48\% for vowel-based naming) prove more challenging, specially for single-instance mining. Misconceptions that require the LLM to reason from the student's perspective (e.g., operator precedence) achieve 42-50\% accuracy. Overall, compared with the single-instance version, multi-instance mining obtains improved performance on more difficult cases.

\section{Conclusion}
\label{sec:conclusion}

This paper introduces {\sc McMiner}, an LLM-based tool for automatically discovering programming misconceptions in student code. Our evaluation on a benchmark of 1,063 code samples exhibiting 67 misconceptions demonstrates strong performance, with the best multi-instance model achieving 82.0\% accuracy. Critically, {\sc McMiner} can identify emerging and rare student misconceptions not captured in existing taxonomies, discovering 41 novel misconceptions in our evaluation. This capability opens the door to uncovering rare, student-specific misconceptions and enables more personalized learning experiences.

Future work includes systematic evaluations of {\sc McMiner} on real student submissions and investigate pedagogical integration in a classroom setting.

\section{Limitations}

Our benchmark assumes each code bag exhibits a single primary misconception, while real students often hold multiple interacting misconceptions simultaneously. The evaluation is limited to Python; generalization to other programming languages remains unexplored. Our misconception dataset contains 67 initial misconceptions and 42 novel misconceptions discovered by LLMs, the largest collection to our knowledge, and is designed to be easily extended, but cannot capture all possible student misconceptions.

% Improving performance is an avenue for future research, particularly, reducing the false positive rate on correct code and its struggle with detecting subtle misconceptions specially those that require code tracing and theory of mind represent important areas for future research to improve precision.

% {\sc McMiner} can only detect misconceptions related to programming that manifest as consistent code patterns. Misconceptions that are not related to programming, such as a student's misinterpreting the problem requirements, or having misconceptions regarding programming concepts like compilers, operating systems, or hardware, are not detectable by {\sc McMiner}.

The LLM-based approach incurs computational costs that may limit large-scale deployment in resource-constrained educational environments. Furthermore, deployment of {\sc McMiner} in real classrooms can raise pedagogical and privacy considerations. Over-reliance on automated detection may reduce instructors' diagnostic skills and discourage valuable Socratic dialogue with students. Instructors using the system may risk premature intervention before students have the opportunity for self-correction through debugging, a critical part of learning. Additionally, processing student code through third-party LLM APIs raises data privacy concerns and requires careful consideration of educational data protection regulations such as FERPA. 

\section*{Acknowledgments}

The Microsoft AFMR program provided generous Azure credits for the LLM experiments. This research was partly supported by the United States Air Force (USAF) under Contract No. FA8750-21-C-0075. Any opinions, findings, conclusions, or recommendations expressed in this material are those of the author(s) and do not necessarily reflect the views of the USAF.

% Bibliography entries for the entire Anthology, followed by custom entries
%\bibliography{anthology,custom}
% Custom bibliography entries only
\bibliography{custom}

\onecolumn
\appendix
\newpage

\section{McInject Qualitative Analysis}
\label{sec:appendix-mcinject-error}

We conducted a qualitative analysis of {\sc McInject}'s generated code samples to understand performance variation across misconception types.

\subsection{Success Patterns}

Several misconceptions achieved 100\% exhibit rates. Analysis reveals these fall into two primary categories:

\noindent{\bf Syntax-based misconceptions} produce clear, unambiguous code patterns that directly violate Python syntax rules. For example, ({\it Student believes that the = operator is used for equality comparison in conditional statements}) and ({\it Student believes that colons (:) are used for variable assignment}) create syntax errors that are straightforward to inject and verify. Similarly,({\it Student believes that a function can be defined without the def keyword}) removes a required keyword, making the misconception immediately apparent.

\noindent{\bf Clear misuse misconceptions} involve well-defined incorrect usage of language constructs where the distinction between correct and incorrect usage is unambiguous. ({\it Student believes that a print statement must be used to return a value from a function}) exemplifies this pattern—replacing {\tt return} with {\tt print} creates a distinct behavioral signature that is both easy to inject and verify. ({\it Student believes that functions are called using square brackets like list indexing}). This misconception similarly produces code with an unmistakable incorrect pattern.

\subsection{Failure Patterns}

Conversely, misconceptions with exhibit rates below 70\% reveal two distinct failure modes that highlight the challenges in generating realistic student-like code.

\subsubsection{Ambiguous Correctness}

Some misconceptions can manifest in code that is both correct and natural, making it difficult to determine whether the code reflects the misconception or simply represents a valid solution approach. Consider the misconception ({\it Student believes that loop iteration requires manual counter tracking with an explicit variable to access element indices}) for the problem {\it ``Write a function to find the next smallest palindrome greater than a given number''} {\sc McInject} generated:

\begin{verbatim}
import sys
def next_smallest_palindrome(num):
    numstr = str(num)
    i = num + 1
    while i < sys.maxsize:
        if str(i) == str(i)[::-1]:
            return i
        i += 1
\end{verbatim}

This code correctly solves the problem and does exhibit manual counter tracking. However, it is difficult to determine whether a student wrote this code due to holding the misconception or simply because this represents a valid and natural solution approach using a while loop. The LLM-as-a-judge in {\sc McInject}'s validation pipeline tends to be conservative in such cases, labeling these samples as not clearly exhibiting the misconception. While this reduces the exhibit rate, it improves dataset quality by ensuring that retained samples exhibit the misconception more clearly.

This pattern particularly affects operator precedence misconceptions, which achieve approximately 50\% exhibit rates. These misconceptions require subtle modifications to arithmetic expressions that can produce correct results or produce unnatural code such as writing {\tt num + 2 * 10} instead of {\tt num + 20}. McInject tends to focus on emulating the code structure rather than writing code that the student with the misconception would write and genuinely believes that it would solve the problem correctly.

\subsubsection{Unnatural Code}

A second failure mode occurs when {\sc McInject} produces valid code matching the misconception pattern but lacking semantic plausibility. Successfully injecting subtle misconceptions requires theory of mind, the LLM must authentically adopt the perspective of a student holding that false belief. When this fails, the tool produces code that exhibits the pattern but makes no logical sense.

For instance, for Misconception 18 ({\it Student believes that in the expression x == a or b, the comparison operator distributes to both operands of the or operator}), {\sc McInject} generated for the same palindrome problem:

\begin{verbatim}
import sys
def next_smallest_palindrome(num):
    numstr = str(num)
    for i in range(num+1, sys.maxsize):
        if str(i) == str(i)[::-1] or str(i)[0]:
            return i
\end{verbatim}

This code is unrealistic. It makes no sense for a student to check whether {\tt str(i) == str(i)[0]} (checking if a string equals its first character) when they could simply check {\tt len(i) == 1}. Furthermore, the or condition is not needed at all. The program is correct without it. The code syntactically matches the misconception pattern but fails the semantic plausibility test.

Similarly, for ({\it Student believes that chained function calls are evaluated from right to left}, {\sc McInject} produced:

\begin{verbatim}
import sys
def next_smallest_palindrome(num):
    numstr = str(num)
    for i in range(num+1, sys.maxsize):
        if str(i).replace('0', '')[::-1] == str(i).replace('0', ''):
            return i
\end{verbatim}

While this attempts to demonstrate order-dependent operations through chained method calls, the code is buggy and nonsensical. Removing zeros from a number before or after reversing is not a correct solution to the palindrome problem and does not reflect a plausible student reasoning path.

\section{McMiner Analysis}
\label{sec:appendix-mcminer-error}

In this section, we examine the performance of McMiner on the benchmark dataset.

\subsection{Performance on Correct-Only vs. Misconception Bags}

A critical evaluation metric for misconception mining tools is their ability to correctly identify when code contains no misconceptions, avoiding false positives that could mislead educators. Analysis of {\sc McMiner-M} performance reveals a gap between these two scenarios across different model configurations, as shown in Table~\ref{tab:mcminer-correct-vs-misc}.

\begin{table}[h]
\centering
\small
\begin{tabular}{@{}lccc@{}}
\toprule
\textbf{Model} & \textbf{Reasoning} & \textbf{Correct-Only} & \textbf{Misc. Bags} \\
 & & \textbf{Accuracy} & \textbf{Accuracy} \\
\midrule
Claude Sonnet 4.5 & \checkmark & 59.30\% & 89.72\% \\
Claude Sonnet 4.5 & $\times$ & 41.86\% & 91.30\% \\
\midrule
OpenAI o3-mini (medium) & -- & 66.28\% & 80.63\% \\
OpenAI o3-mini (low) & -- & 75.58\% & 75.49\% \\
\midrule
Gemini 2.5 Flash & \checkmark & 36.05\% & 90.91\% \\
Gemini 2.5 Flash & $\times$ & 61.63\% & 60.87\% \\
\bottomrule
\end{tabular}
\caption{Performance comparison of {\sc McMiner-M} across different models on correct-only bags versus bags containing misconceptions. All metrics are based on the 339 total bags (279 bags with misconceptions and 60 correct-only bags).}
\label{tab:mcminer-correct-vs-misc}
\end{table}

The results reveal substantial variation in false positive rates across models. Claude Sonnet 4.5 with reasoning achieves 89.72\% accuracy on misconception bags but only 59.30\% on correct-only bags. Without reasoning, this gap widens (91.30\% vs. 41.86\%), indicating that reasoning capabilities help reduce false positives. OpenAI o3-mini (low) demonstrates the best balance with nearly equal performance on both bag types (75.58\% vs. 75.49\%).

\subsubsection{Common False Positive Patterns}

Analysis of false positives on correct-only bags reveals systematic patterns in the types of ``misconceptions'' models incorrectly identify {\bf Inefficiencies as Misconceptions}. This is where models sometimes identify ``inefficiencies'' as misconceptions, such as flagging manual iteration instead of using built-in functions like {\tt enumerate()} or list comprehensions. While these may represent less idiomatic code, they do not constitute misconceptions about language semantics.

\subsection{Novel Misconception Discovery}

A key strength of {\sc McMiner} is its ability to discover misconceptions not present in the predefined misconception bank. Using LLM-as-a-judge validation (Section~\ref{sec:appendix-evaluation-prompts}), we identified novel true positives—predicted misconceptions that, while not matching ground truth, accurately describe genuine programming misunderstandings exhibited in the code.

Claude Sonnet 4.5 with reasoning discovered 41 novel true positives across the 339 bags (12.1\%), while OpenAI o3-mini (medium effort) found 35 (10.3\%).

\subsubsection{Cherry-Picked Novel Discovery}

\noindent{\bf Example: Division Operator Type Semantics}

For a bag containing code implementing tetrahedral number (which is always an integer) calculation, {\sc McMiner} identified:

\begin{quote}
\small
\noindent{\bf Predicted Misconception:} ``The student believes that the division operator {\tt /} and integer division operator {\tt //} are interchangeable or that {\tt /} will automatically return an integer when the result is a whole number.''

\noindent{\bf Code Context:}
\begin{verbatim}
def tetrahedral_number(n): 
    return (n * (n + 1) * (n + 2)) / 6
\end{verbatim}

\noindent{\bf Explanation:} In Python 3, the {\tt /} operator always returns a float, even when dividing two integers that result in a whole number. For calculating tetrahedral numbers, which are always integers, using {\tt //} would be more appropriate to return an integer type.
\end{quote}

This misconception was not in the original bank but represents a valid and common misunderstanding about Python 3's division semantics.

\subsection{Error Analysis}

Certain misconceptions in the benchmark proved particularly challenging for {\sc McMiner}, with several achieving near-zero detection rates. Analysis reveals these share a common characteristic: they are mostly {\it benign misconceptions}, false beliefs that do not cause functional bugs but instead reflect stylistic preferences or suboptimal practices.

\noindent{\bf Low-Performing Misconceptions:}
\begin{itemize}[leftmargin=*,itemsep=2pt,topsep=3pt]
\item ``Student believes that the {\tt return} statement requires parentheses around its argument'' code like {\tt return(x + y)} is syntactically valid and functions correctly
\item ``Student believes that function parameters automatically change in recursive calls without explicit modification.''. This is a harmful misconception that can cause infinite recursion.
\item ``Student believes that loop iteration requires manual counter tracking with an explicit variable to access element indices.''. This benign misconception can often result in functionally correct code that is also natural and lacks unusual features.
\end{itemize}

These benign misconceptions fail detection because:

\begin{enumerate}[leftmargin=*,itemsep=2pt,topsep=3pt]
\item {\bf Code is functionally correct}: Models trained on code understanding tasks prioritize functional correctness, making them less sensitive to stylistic variations
\item {\bf Patterns lack distinctive signatures}: Unlike misconceptions that produce syntax errors or logical bugs, benign misconceptions produce code indistinguishable from intentional stylistic choices
\item {\bf Ambiguous intent}: Without access to the student's reasoning, it is impossible to definitively determine whether {\tt return(x)} reflects a misconception or a stylistic preference influenced by other programming languages
\end{enumerate}

\section{McMiner Interface}
\label{sec:appendix-interface}
To facilitate the use of McMiner for educators and researchers, we developed an interactive web application using the Streamlit\footnote{\url{https://streamlit.io}} Python library that implements the McMiner-S single-instance mining approach. The interface allows users to input a problem description and student code, then select from multiple state-of-the-art LLMs (Claude, GPT, Gemini) to analyze the code for potential programming misconceptions. The tool uses the same prompt template and model configurations as our benchmark experiments, automatically enabling reasoning capabilities for compatible models. Results are displayed with structured misconception descriptions, explanations of how they manifest in the code, and expandable reasoning traces from the model. The application loads user API credentials securely from environment variables and provides real-time analysis of student code within seconds.

\begin{figure}[!h]
    \centering
    \includegraphics[width=\linewidth]{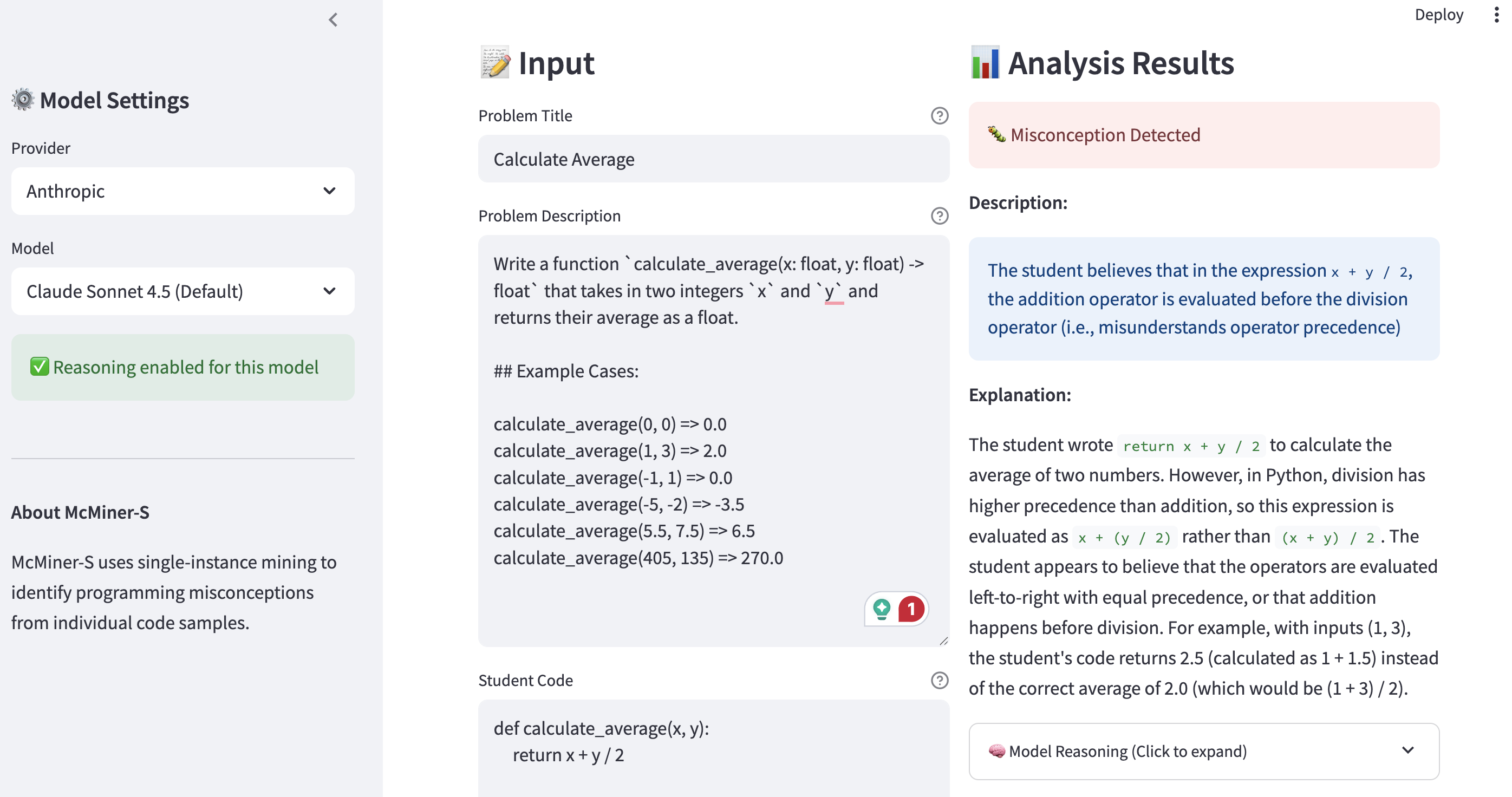}
    \caption{Web application interface of the McMiner-S tool. Users are able to input a programming problem and a student implementation, and analyze the code for any potential misconceptions.}
    \label{fig:mcminer-interface}
\end{figure}

% \section{Qualitative Analysis of McMiner on Real Student Submissions}
% \label{sec:appendix-qualitative}

% \tcr{TODO.}

\section{Dataset Sources}
\label{sec:appendix-dataset}

Our misconceptions were adapted from: The Python Misconception bank~\cite{python_misc_bank_2021} (17 misconceptions), SIDELib~\cite{sidelib_2023} (36 misconceptions), and 14 were composed by the authors, 3 of which are out-of-domain to test the ability of misconception mining tools to identify previously unseen patterns. The three out-of-domain misconceptions are: (1) {\it Student believes that every variable must be explicitly deleted with 'del' after its last use to prevent memory leaks}, (2) {\it Student believes that variable names containing vowels (a, e, i, o, u) can only store string values, while consonant-only names can store any type}, and (3) {\it Student believes that list indexing starts at -1}. While we have 67 misconceptions, our benchmark contains 1,063  code samples exhibiting those misconceptions. Our benchmark can be easily extended given any new misconception.

The programming problems, solutions, and unit tests in our benchmark dataset were curated from multiple established sources to ensure diversity and representativeness. Each problem includes comprehensive test cases, with an average of 3.31 unit tests per problem to validate correctness. The majority of problems (438 out of 501, or 87.4\%) were adapted from the MBPP dataset~\cite{austin2021program}, which provides a comprehensive collection of simple computational problems in Python with corresponding solutions and test cases. To supplement this core set, we incorporated 27 problems from the Socratic Debugging dataset~\cite{erfan2023_bea,erfan_2024_sigcse}, which focuses on debugging scenarios, along with 8 problems from the Auckland dataset~\cite{ettles_ace_2019} and 6 from the FalconCode dataset~\cite{falconcode_2023}. Finally, 19 problems were handwritten by the authors, drawing inspiration from public programming courses and educational websites such as w3resource to fill gaps in programming language construct coverage.

Table~\ref{tab:benchmark-stats} summarizes the key statistics of our benchmark dataset. The dataset was generated from 67 distinct misconceptions applied to 25 randomly sampled problem-solution pairs ($N_k = 25$) from our curated set of 501 problems. The {\sc McInject} tool generated 1,675 code samples, of which 498 were deemed inapplicable due to incompatibility between the misconception and the problem-solution pair, resulting in 1,177 generated code samples. These samples were then validated using LLM-as-a-judge evaluation, which filtered out 114 samples that did not properly exhibit their intended misconceptions. This yields 1,063 code samples that successfully exhibit misconceptions. Code samples that were deemed inapplicable or did not properly exhibit their intended misconceptions were replaced with the original correct solution and have a null misconception. This yields a final dataset of 1,675 code samples. To enable evaluation in a Multiple Instance Learning setting (later described in Section~\ref{sec:mcminer}), we organized these samples into 339 bags, each containing 4-8 code samples. This bag structure allows us to evaluate both single-instance mining ({\sc McMiner-S}) and multi-instance mining ({\sc McMiner-M}). Importantly, 60 bags (17\% of the bags) contain only correct code samples without any misconceptions, serving as negative examples to test the system's ability to correctly identify when no misconceptions are present.

\subsection{Misconception Description Writing Guidelines}
\label{sec:appendix-misconception-guidelines}

All misconceptions (composed, adapted, and ooD) were written by hand and include one example of a code snippet exhibiting that misconception. Figure~\ref{fig:misconception-guidelines} shows the writing guidelines used to create the misconception bank.

\begin{figure}[!h]
\begin{newshaded}
\small
\noindent {\bf Operational definition}: A concise, one-sentence statement of the student's incorrect belief about a programming construct, operation, or API, excluding the correct concept.

\vspace{0.5em}
\noindent {\bf Guidelines}
\begin{itemize}[leftmargin=*,itemsep=0pt,topsep=3pt]
\item {\bf State ONLY the incorrect belief}: Use ``Student believes that...'' without mentioning or implying the correct construct, concept, or fix.
\item {\bf Avoid line numbers}: Reference only the operation or built-in/API involved, not specific line numbers.
\item {\bf Be concrete and specific about the construct}: Mention the relevant operation, method, or construct (e.g., {\tt list.pop}, string indexing, {\tt range} bounds, conditional statements).
\item {\bf Focus on ONLY one construct}: Write the misconception in terms of only one construct, operation, or API. Avoid general misconceptions applicable to multiple constructs.
\item {\bf Generalize the misconception}: The misconception should be applicable to multiple implementations, not specific to a single problem or implementation.
\item {\bf Exclude inputs, outputs, and fixes}: Do not include example inputs, expected/actual outputs, or the bug fix.
\item {\bf Keep it concise}: One clear sentence that maps directly to what the student implemented.
\end{itemize}
\end{newshaded}
\caption{Misconception description writing guidelines.}
\label{fig:misconception-guidelines}
\end{figure}

\section{The McInject Tool}
\label{sec:appendix-mcinject}

The {\sc McInject} tool leverages Claude-Sonnet-4.5 with extended thinking enabled to implement the misconception injection task. We use temperature 1.0, max\_tokens 6000 (4000 output + 2000 thinking budget), and a 2000 token thinking budget. The tool uses zero-shot prompting with structured XML output format, where the LLM is instructed to modify the correct solution such that it appears as if written by a student who genuinely believes their approach is correct, despite holding the specified misconception.

To enhance generation quality, the prompt includes illustrative examples alongside misconception descriptions. Each misconception in the dataset includes a concrete example demonstrating how the false belief manifests in code, which helps guide {\sc McInject} to produce more realistic student-like code. The McInject prompt template is shown in Figure~\ref{fig:prompt-mcinject}.

To further improve code quality, we implement an iterative refinement process. After {\sc McInject} generates code, an LLM-as-a-judge evaluates whether the code properly exhibits the target misconception (see Section~\ref{sec:appendix-mcinject-judge}). If the judge provides actionable feedback indicating the misconception is not clearly exhibited, this feedback is appended to the original conversation context, and {\sc McInject} is prompted to refine the code. This feedback loop can iterates once.

\begin{figure}[!h]
\begin{newshaded}
\small

\noindent {\bf Your Task}

You will be given as input a programming *problem*, *code* implementing a solution, and the description of a *misconception*. A misconception is a false belief that the student holds about some programming language construct. Note that, misconceptions do not always result in buggy code. Some misconceptions lead to stylistic differences or inefficiencies rather than errors. For example, a student who believes "all variable identifiers must use only one letter" might write working but less readable code.

Modify the code such that it looks as if written by a student who has that misconception. The student genuinely believes that this modified code solves the problem correctly.

\vspace{0.5em}
\noindent {\bf Input Format}

\textbf{Problem Description:} [problem\_description]

\textbf{Given Implementation:}
\begin{verbatim}
[correct_solution]
\end{verbatim}

\textbf{Misconception Description:} 
\begin{verbatim}
[misconception_description]

**Example:**
[misconception_example]
\end{verbatim}

\vspace{0.5em}
\noindent {\bf Output Format}

\begin{verbatim}
<code>
[The complete modified Python code exhibiting the 
misconception]
</code>
\end{verbatim}

If the misconception relates to language constructs that are not present in the given solution, output:
\begin{verbatim}
<code>
NONE
</code>
\end{verbatim}
\end{newshaded}
\caption{Prompt template for {\sc McInject} tool.}
\label{fig:prompt-mcinject}
\end{figure}

A key feature of {\sc McInject} is its ability to handle incompatible cases where a misconception cannot be meaningfully applied to a given solution. Misconceptions are considered incompatible when they relate to language constructs or concepts that are not present or relevant in the solution. For example, a misconception about loop behavior would be incompatible with a solution that contains no loops. In such cases, the tool indicates inapplicability rather than forcing inappropriate modifications.

All code samples produced by {\sc McInject} undergo post-processing where inline comments are automatically removed to ensure clean output. This process uses Python's tokenize module and includes syntax validation to maintain code correctness.

To validate the quality of generated code samples by McInject, we employed LLM-as-a-Judge evaluation showing that 90.3\% of the code samples (1,063 out of 1,177 generated samples) successfully exhibit their target misconceptions. We also employ manual evaluation of 38 samples composed of a misconception and a problem-solution pair that were deemed to be inapplicable by McInject, and observe 100\% agreement with human judgment.

We note that this is a conservative estimate of McInject's performance, since the LLM-as-a-judge discards code samples that do exhibit the misconception, yet are both correct and natural, making the misconception extremely difficult to detect. Manual evaluation of 88 code samples demonstrates 96.6\% agreement between LLM judgment and human assessment, confirming the reliability of the LLM-as-a-judge approach in determining whether a code sample exhibits a misconception description. The LLM-as-a-judge leveraged Claude-Sonnet-4.5 with the same hyperparameters as McInject.

% \vspace{-1cm}

\section{The McMiner Tools}
This section lists all the prompts that are used by the McMiner tools.

\label{sec:appendix-mcminer}

\begin{figure}[!h]
\begin{newshaded}
\small
\noindent {\bf Key Terminology}

\vspace{0.5em}
A {\bf programming misconception} refers to a false belief that a student holds about some programming language construct or built-in function in Python. Programming misconceptions can be about the syntax or the semantics of constructs in the Python programming language. They should not be about concepts in the problem description.

\vspace{0.5em}
\noindent {\bf Your Task}

Given a problem description and student code that attempts to solve that problem, identify a most likely programming misconception that is exhibited by that code, if any.

\vspace{0.5em}
\noindent {\bf Input Format}

Problem Description: \{problem\_description\}

Student Code:
\begin{verbatim}
{student_code}
\end{verbatim}

\vspace{0.5em}
\noindent {\bf Output Format}

\begin{verbatim}
<misconception>
<description>[Describe misconception, starting
with "The student believes"]</description>
<explanation>[Explain how the code exhibits
             the misconception]</explanation>
</misconception>
\end{verbatim}

If no misconceptions are found, output:
\begin{verbatim}
<misconception>NONE</misconception>
\end{verbatim}
\end{newshaded}
\caption{Prompt template for {\sc McMiner-S} (single-instance mining). The full template includes additional metadata fields and guidelines.}
\label{fig:prompt-mcminer-s}
\end{figure}

\begin{figure}[!h]
\begin{newshaded}
\small
\noindent {\bf Key Terminology}

\vspace{0.5em}
A {\bf programming misconception} refers to a false belief that a student holds about some programming language construct or built-in function in Python. Programming misconceptions can be about the syntax or the semantics of constructs in the Python programming language. They should not be about concepts in the problem description. For example, "Student believes range(n) produces values from 1 to n inclusive" is a valid programming misconception about Python's range() function, while "Student thinks natural numbers can be negative" is not a programming misconception. Also, while misconceptions often cause bugs, sometimes they do not. For example, "Student believes all variable names need to have exactly one letter" is a programming misconception that does not necessarily cause a bug.

\vspace{0.5em}
\noindent {\bf Your Task}

Input: a set of <problem description, code> pairs, where the code in each pair attempts to solve the corresponding problem. 
Output: the description of a programming misconception that is exhibited by one or more code samples in the input set. While the input set will often contain code samples that show no misconception, try to identify a misconception that is exhibited by most code samples in the input set. If you cannot find any misconception, output NONE.

\vspace{0.5em}
\noindent {\bf Input Format}

Problem Description: \{problem\_description\}

Student Code:
\begin{verbatim}
{student_code}
\end{verbatim}

\vspace{0.5em}
\noindent {\bf Output Format}

\begin{verbatim}
<misconception>
<description>[Describe misconception, starting
with "The student believes"]</description>
<explanation>[Explain how the code exhibits
             the misconception]</explanation>
</misconception>
\end{verbatim}

If no misconceptions are found, output:
\begin{verbatim}
<misconception>NONE</misconception>
\end{verbatim}
\end{newshaded}
\caption{Prompt template for {\sc McMiner-M} (multi-instance mining). The full template includes additional metadata fields and guidelines.}
\label{fig:prompt-mcminer-m}
\end{figure}

\section{Experimental Setup}
\label{sec:appendix-experimental}

We evaluated {\sc McMiner} using three state-of-the-art LLMs through their respective APIs:

\begin{itemize}
    \item {\bf OpenAI}: We used the o3-mini model with two reasoning effort levels. For low effort, we set reasoning\_effort to "low" and max\_completion\_tokens to 7000 (4000 output + 3000 reasoning). For medium effort, we set reasoning\_effort to "medium" and max\_completion\_tokens to 9000 (4000 output + 5000 reasoning). Note that o-series models do not support temperature configuration.
    
    \item {\bf Anthropic}: We used the claude-sonnet-4-5 model. For non-reasoning experiments, we used temperature 0.1 and max\_tokens 4000. For reasoning-enabled experiments, we used temperature 1.0 (as required by Anthropic for extended thinking), max\_tokens 6000 (4000 output + 2000 thinking budget), and enabled extended thinking with a 2000 token budget.
    
    \item {\bf Google Gemini}: We used the gemini-2.5-flash model with temperature 0.1 and max\_tokens 4000. For reasoning-enabled experiments, we increased max\_tokens to 6000 (4000 output + 2000 thinking budget) and enabled thinking with a 2000 token budget.
\end{itemize}

All experiments used individual request processing (no batching) with a zero-shot prompting strategy. For evaluation, we used the benchmark dataset $\mathcal{D}$ described in Section 4, containing the 339 bags of code samples. 

% For {\sc McMiner-S}, we processed each problem-code pair in $PC^{(k)}$ individually, resulting in 1,375 predictions. For {\sc McMiner-M}, we used the 339 bags of code samples as described in Section~\ref{sec:multiple}.

To evaluate whether predicted misconceptions semantically match ground truth (GT) misconceptions, we employed an LLM-as-judge approach using Claude Sonnet-4.5 with reasoning capabilities. The evaluation uses identical hyperparameters as {\sc McInject}. Claude evaluates whether the predicted misconception description matches the intended misconception.

% accounting for variations in wording while focusing on the fundamental programming misunderstanding.

To credit novel misconception discoveries not present in our benchmark, we implement a novelty-aware evaluation approach. For predictions that do not match GT, we use the same LLM-as-judge validation from {\sc McInject} to determine whether the code samples accurately exhibit the predicted misconception. If validated, these are counted as novel true positives.

\section{Evaluation Prompts}
\label{sec:appendix-evaluation-prompts}

This section discusses all the prompts used for evaluation in this paper.

\subsection{McInject LLM-as-a-judge Prompt}
\label{sec:appendix-mcinject-judge}

To validate that code samples generated by {\sc McInject} properly exhibit their intended misconceptions, we employ an LLM-as-a-judge evaluation. The judge is given a misconception description with an example, and a code sample to analyze. The task is to determine whether the code exhibits the misconception, providing a binary judgment (Y/N) and optional feedback. 

When used in the iterative refinement process, the feedback mechanism works as follows: if the judge determines the misconception is not clearly exhibited and provides specific feedback (i.e., feedback is not ``NONE''), this feedback is used to create a multi-turn conversation. The original {\sc McInject} prompt and its initial response are preserved, and the feedback is appended as a new user message instructing {\sc McInject} to improve the code based on the critique. This creates a conversational refinement loop where {\sc McInject} can iteratively improve its output. The process terminates when either the judge confirms the misconception is exhibited (feedback is ``NONE''), or the iteration limit is reached.

Importantly, the prompt emphasizes that misconceptions do not necessarily cause bugs—code can be syntactically and logically correct while still exhibiting a false belief pattern. Figure~\ref{fig:prompt-mcinject-judge} shows the prompt template used for this evaluation.

\begin{figure}[!h]
\begin{newshaded}
\small
\noindent {\bf Input Format}

\noindent {\bf The Misconception}
\begin{verbatim}
<misconception>
Description: {misconception_description}
Example: {misconception_example}
</misconception>
\end{verbatim}

\noindent {\bf The Code to Analyze}
\begin{verbatim}
<code>
{code_to_analyze}
</code>
\end{verbatim}

\vspace{0.5em}
\noindent {\bf Your Task}

Determine whether this code exhibits the misconception described above.

\vspace{0.5em}
\noindent {\bf Key Understanding}

{\bf A misconception does NOT necessarily induce bugs or errors!} Code can be:
\begin{itemize}[leftmargin=*,itemsep=0pt,topsep=3pt]
\item {\bf Syntactically correct} (no syntax errors)
\item {\bf Logically correct} (produces expected output)
\item {\bf Yet still exhibit a misconception} (shows the student holds a false belief)
\end{itemize}

\vspace{0.5em}
\noindent {\bf Analysis Guidelines}
\begin{enumerate}[leftmargin=*,itemsep=2pt,topsep=3pt]
\item {\bf Understand the misconception deeply}: What incorrect belief does the student have? What coding patterns would reveal this belief?
\item {\bf Analyze the code systematically}: Look for patterns that match the misconception. Check if the code structure reflects the incorrect belief.
\item {\bf Focus on the belief, not the outcome}: Does the code structure suggest the student holds this false belief? Even if the code works, does it show the misconception pattern?
\end{enumerate}

\vspace{0.5em}
\noindent {\bf Output Format}
\begin{verbatim}
<answer>
<exhibits_misconception>Y or N</exhibits_misconception>
<feedback>[Optional feedback]</feedback>
</answer>
\end{verbatim}
\end{newshaded}
\caption{Prompt template for LLM-as-a-judge evaluation of {\sc McInject} generated code samples. The full prompt template includes metadata fields to output such as rationale and confidence level.}
\label{fig:prompt-mcinject-judge}
\end{figure}

\subsection{McMiner Semantic Matching Prompt}
\label{sec:appendix-mcminer-judge}

To evaluate {\sc McMiner} predictions, we use an LLM-as-a-judge to determine whether a predicted misconception semantically matches the ground truth misconception. The judge is provided with the ground truth misconception, the predicted misconception, and the code samples that were analyzed. The task is to assess whether both descriptions capture the same fundamental programming misunderstanding, accounting for natural variations in wording. The evaluation considers core concept alignment, evidence in the code samples, and semantic equivalence. Figure~\ref{fig:prompt-semantic-match} shows the prompt template used for this evaluation.

\begin{figure}[!h]
\begin{newshaded}
\small
\noindent {\bf Input Format}

\noindent {\bf Ground Truth Misconception}
\begin{verbatim}
{ground_truth}
\end{verbatim}

\noindent {\bf Predicted Misconception}
\begin{verbatim}
{predicted_misconception}
\end{verbatim}

\noindent {\bf Code Samples Analyzed}
\begin{verbatim}
{code_samples}
\end{verbatim}

\vspace{0.5em}
\noindent {\bf Task}

Determine if the predicted misconception accurately captures the same conceptual misunderstanding as the ground truth. Consider:

\begin{enumerate}[leftmargin=*,itemsep=2pt,topsep=3pt]
\item {\bf Core Concept Match}: Are they describing the same fundamental misunderstanding about programming concepts?
\item {\bf Evidence Alignment}: Does the predicted description align with what's shown in the code samples?
\item {\bf Semantic Equivalence}: Minor wording differences are acceptable if the core concept matches.
\item {\bf Example Reference}: If provided, use the misconception example to better understand the typical manifestation of this misconception.
\end{enumerate}

\vspace{0.5em}
\noindent {\bf Special Cases}
\begin{itemize}[leftmargin=*,itemsep=0pt,topsep=3pt]
\item If ground truth is ``NO MISCONCEPTION'' and prediction found no misconceptions, this is a match.
\item If ground truth is ``NO MISCONCEPTION'' but prediction found misconceptions, this is NOT a match.
\item If ground truth describes a misconception but prediction found none, this is NOT a match.
\end{itemize}

\vspace{0.5em}
\noindent {\bf Output Format}
\begin{verbatim}
<evaluation>
<match>true or false</match>
</evaluation>
\end{verbatim}
\end{newshaded}
\caption{Prompt template for semantic matching evaluation of {\sc McMiner} predictions. The full prompt template includes metadata fields to output such as confidence level and explanation.}
\label{fig:prompt-semantic-match}
\end{figure}

% \section

% \subsubsection{Standard Evaluation Metrics}

% Accuracy represents the proportion of correctly predicted bag labels. For 
% precision, recall, and F1-score: \textbf{True Positives (TP)} are cases where 
% ground truth contains a misconception that matches the prediction; \textbf{True 
% Negatives (TN)} are cases where both ground truth and prediction contain no 
% misconception; \textbf{False Positives (FP)} are cases where the model predicted 
% a misconception but ground truth contains none, or the predicted misconception 
% doesn't match ground truth; \textbf{False Negatives (FN)} are cases where ground 
% truth contains a misconception but the model predicted none.

\end{document}